\documentclass{article}

     \usepackage[preprint]{neurips_2021}

\usepackage[utf8]{inputenc} 
\usepackage[T1]{fontenc}    
\usepackage{hyperref}       
\usepackage{url}            
\usepackage{booktabs}       
\usepackage{amsfonts}       
\usepackage{nicefrac}       
\usepackage{microtype}      
\usepackage{xcolor}         

\usepackage{graphicx}
\usepackage{multicol}
\usepackage{multirow}
\usepackage{graphics} 
\usepackage{epsfig} 
\usepackage{amsmath} 
\usepackage{amssymb}
\usepackage[small]{caption}
\usepackage[export]{adjustbox}
\usepackage{subcaption}

\usepackage{algorithm}
\usepackage{algorithmic}

\newcommand{\kalesha}[1]{{\textcolor[rgb]{0.75,0.0,0.75}{KB: #1}}}

\newcommand{\ourmethod}{QED}

\long\def\|*#1*/{}

\DeclareMathOperator*{\argmax}{arg\,max}

\title{Quasi-Equivalence Discovery for \\Zero-Shot Emergent Communication}

\author{Kalesha Bullard \\
	Facebook AI Research \\
	\texttt{ksbullard@fb.com} \\
	\And
	Douwe Kiela \\
	Facebook AI Research \\
	\texttt{dkiela@fb.com} \\
	\And
	Franziska Meier \\
	Facebook AI Research \\
	\texttt{fmeier@fb.com} \\
	\And
	Joelle Pineau \\
	Facebook AI Research, MILA \\
	\texttt{jpineau@fb.com} \\
	\And
	Jakob Foerster \\
	Facebook AI Research \\
	\texttt{jnf@fb.com} \\
}

\|*
\author{%
  David S.~Hippocampus\thanks{Use footnote for providing further information
    about author (webpage, alternative address)---\emph{not} for acknowledging
    funding agencies.} \\
  Department of Computer Science\\
  Cranberry-Lemon University\\
  Pittsburgh, PA 15213 \\
  \texttt{hippo@cs.cranberry-lemon.edu} \\
} */

\begin{document}

\maketitle

\begin{abstract}


Effective communication is an important skill for enabling information exchange in multi-agent settings and \emph{emergent communication} is now a vibrant field of research, with common settings involving discrete cheap-talk channels. Since, by definition, these settings involve arbitrary encoding of information, typically they do not allow for the learned protocols to generalize beyond training partners. 
In contrast, in this work, we present a novel problem setting and the Quasi-Equivalence Discovery (\ourmethod) algorithm that allows for \emph{zero-shot coordination} (ZSC), \emph{i.e.}, discovering protocols that can generalize to independently trained agents. Real world problem settings often contain \textit{costly} communication channels, \emph{e.g.}, robots have to physically move their limbs, and a non-uniform distribution over intents. We show that these two factors lead to unique optimal ZSC policies in referential games, where agents use the energy cost of the messages to communicate intent. \emph{Other-play} was recently introduced for learning optimal ZSC policies, but requires prior access to the symmetries of the problem. Instead, \ourmethod{} can iteratively \emph{discovers} the symmetries in this setting and converges to the optimal ZSC policy. 
 



\end{abstract}


\section{Introduction}
	
\iftrue
The ability to communicate effectively with other agents is part of a necessary skill repertoire of intelligent agents and commonly seen as one of the great achievements of humanity.  
%
A number of papers have studied \emph{emergent communication} in multi-agent settings~\citep{foerster2016learning,lazaridou2016multi,lowe2019interaction}~\citep[for a review, see][]{lazaridou2020emergent}. The most common problem settings studied are communication tasks, in which agents can exchange messages through symbolic (discrete) cheap-talk channels, that have no impact on the reward function or transition dynamics. A common task is the so called \emph{referential game}, in which a \emph{sender} observes an \emph{intent} that needs to be communicated to a \emph{listener} via a \emph{message}.

\nocite{lowe2019pitfalls,evtimova2018emergent,lazaridou2018emergence,kottur2017natural}

In these cheap-talk settings, the solution space typically contains many equivalent but mutually incompatible (self-play) policies. For example, permuting bits in the channel and adapting the receiver policy accordingly would preserve payouts, but differently permuted senders and receivers are mutually incompatible. This makes it difficult for independently trained agents to utilize the cheap-talk channel at test time, a setting formalized as zero-shot coordination~\citep[ZSC;][]{hu2020other}. 

\begin{figure*}[t]
 	\centering
 	\includegraphics[width=0.75\textwidth]{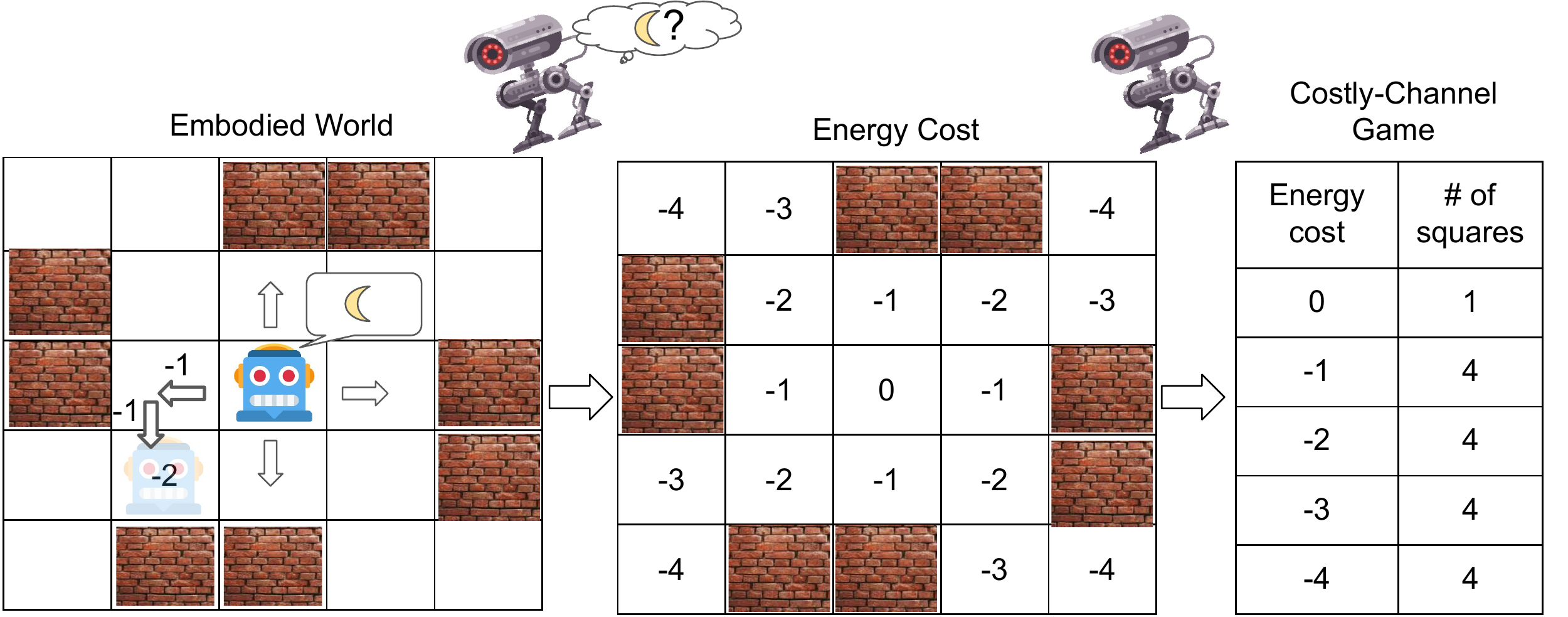}
 	\caption{\textbf{Costly-Channel Referential Game.} Sender agent (robot) must take \textit{costly} physical actions to move to target locations on the grid.  The energy cost incurred corresponds to the number of steps traversed.  Receiver agent (camera) observes final location of sender and must predict what intent the action taken \textit{implicitly} communicates.}  
 	\label{fig:embodied_referential_game}
 	\vspace{-4mm}
\end{figure*}

In contrast, we present a minimum viable setting, inspired by real-world problems, in which zero-shot communication is \emph{in principle} possible. Specifically, our setting deviates from the standard referential game in two ways: First of all, we assume that communication actions are \emph{costly}, which is inspired by the fact that agents in the real world commonly have to communicate through their actions. As a specific example, imagine a robot that needs to communicate with another agent by moving its limbs. Clearly, this actuation will require the expense of energy and different actions will require different amounts of energy. Since our message cost is a proxy for this energy we will refer to our cost as \emph{energy} in the rest of the paper (obviously there could be other common knowledge costs).  

Secondly, another ubiquitous factor of the physical world (and many other domains) is that communicative intents are \textit{not} distributed uniformly. In particular, the Zipf distribution~\citep{zipf2016human} is known to be a good proxy for many real-world distributions associated with human activity.

Combining a variable message cost with a non-uniform intent distribution over intents, in the context of referential games, allows for ZS Communication:
Actions requiring lower energy exertion should be used to encode more common intents, while those requiring higher energy encode less common ones. As we illustrate in Section~\ref{sec:experiments}, this is the unique solution to our ZS communication problem.

Auxiliary losses, such as entropy penalties, are superficially related but do not allow for ZSC without further assumptions. While these auxiliary losses are design decisions, energy cost is an example of a universal  (common-knowledge) cost, grounded in the environment, which can be exploited for ZSC.  

Unfortunately, training agents that can successfully learn ZSC strategies is a difficult problem for current machine learning models. There are two major challenges: 
1) The latent structure underlying the protocol, in our case energy, that can be coordinated on, needs to be discovered, requiring agents to ignore other (redundant) degrees of freedom. 
2) Local optima associated with the lock-in between sender and receiver can trap the learning algorithm, since the interpretation of a message depends on the entire policy, not just the state and action. 

The main challenge associated with (1) is due to \emph{energy degeneracy}: When different actions have the same energy cost, optimal self-play policies will arbitrarily break the symmetry and use each of the actions to encode a different intent. Clearly, this is not an option for ZS communication. To address this kind of coordination failure, recently \emph{Other-Play} (OP) \cite{hu2020other} was introduced. OP assumes that the symmetries of a problem setting are provided ahead of time and uses them to define an expected return over equivalence classes of policies, rather than specific policies. Unfortunately, as pointed out by the authors, learning the symmetries for a Dec-POMDP is in general a hard problem.

In this work, we introduce the Quasi-Equivalence Discovery (\ourmethod) algorithm which seeks to recover equivalence mappings from a set of optimal \emph{self-play} policies. \ourmethod{} then uses those equivalence mappings in combination with the other-play algorithm to train a new set of policies from which further equivalences can be recovered. 

We evaluate the \ourmethod{} algorithm on a referential game with costly actions, with and without energy degeneracy. 
As we show, \ourmethod{} converges to the unique ZSC policy, recovering the performance of other-play with ground truth access to the symmetries, and this presents an important step towards developing machine learning methods that can coordinate with humans and other agents. 

Perhaps one of the most valuable contributions of this paper is to shed light on the \textit{regularization} mechanisms that can be effective in ZSC, namely cost/energy and non-uniform/Zipf distribution. 


\fi

\section{Related Work}
\label{sec:related}

\iftrue

Emergent communication in multi-agent settings has been an area of active study in recent years \citep{foerster2016learning,lazaridou2016multi,havrylov2017emergence,cao2018emergent,bouchacourt2018agents,eccles2019biases,graesser2019emergent,chaabouni2019anti,jaques2019social,lowe2019interaction}. These papers typically assume a symbolic (discrete) cheap-talk channel, through which agents can send messages that have no impact on the reward function or transition dynamics. A common task is the so called \emph{referential game}, in which a \emph{sender} observes an \emph{intent} needing to be communicated to a \emph{listener} via a \emph{message}.
As a main difference, our problem setting is \emph{zero-shot coordination}, rather than \emph{self-play}, which is feasible since our messages are \emph{costly} and the distribution over intents \emph{non-uniform}.
\nocite{lowe2019pitfalls,evtimova2018emergent,lazaridou2018emergence,kottur2017natural}

There has been work on the emergence of grounded language for robots \citep{steels2012emergent1,spranger2016evolution}. Like in the emergent communication work mentioned above, the goal of this research is to develop self-play policies.  To the best of our knowledge, we are the first to develop methods that can accomplish \emph{zero-shot emergent communication}.  

\nocite{steels2001language,steels2012emergent2,spranger2012emergent}

There have also been a number of different approaches for moving beyond the \emph{self-play} setting. \emph{Ad-Hoc Teamwork}~\citep{stone2010ad}, assumes that there is a given population of teams and trains agents to do well when substituted-in as one of the members of these teams.  In contrast, in the \emph{zero-shot coordination} setting, the task is to find a self-consistent learning algorithm that allows independently trained agents to coordinate on the first attempt. Thus, crucially, we do not assume that a population of teams is provided along with the problem setting. 
The question of how agents can coordinate on a given problem instance has also been studied extensively in the fields of game theory, economics and behavioural psychology. Common approaches include the \emph{cognitive hierarchies model} \cite{camerer2004cognitive} and the \emph{focal points theory} \cite{schelling1980strategy}. Clearly, both cognitive hierarchies  and focal point approaches will fail entirely in our setting. Cognitive hierarchies can never learn any signalling behavior, since our messages have no \emph{grounded meaning} and focal points are not applicable since we assume the different parties do not share any common labels for states and actions (as is typical in the ZSC setting). 

Another related body of work is the research on linguistics and pragmatic reasoning in language, \emph{e.g.} the seminal work by ~\cite{frank2012predicting}. Unfortunately, just like the cognitive hierachy approaches mentioned, this is not applicable in our setting since the messages have no a-priori grounded meaning. As such, a naive speaker is entirely uninformative and can \emph{never} induce informative posteriors, even in higher levels. 

The closest to our method is the \emph{other-play} method from~\cite{hu2020other} that we build upon. The two crucial differences are that (1) our approach can \emph{discover} the symmetries through learning rather than requiring them to be provided with the Dec-POMDP and that (2) we investigate an emergent communication setting in which the communication actions have no prior grounded meaning.
\fi
	

\section{Background}
\label{sec:related}

\iftrue

\subsection{Multi-Agent Reinforcement Learning}
We formalize the protocol learning problem as a decentralized partially observable Markov decision process (Dec-POMDP) with $N$ agents \citep{bernstein2002complexity}, defined by tuple $(S,A,T,R,O,\mathbb{O},\gamma)$.  The Dec-POMDP model extends a single-agent POMDP by considering \textit{joint} actions and observations.  $S$ is the set of states, $A = A_1 \cdot \cdot \cdot A_N$ the set of actions for each of $N$ agents in the population (the joint action), and $T$ a transition function $S \times A_1 \cdot \cdot \cdot A_N \rightarrow S$ mapping each state and joint action to a distribution over next states.  In a partially observable setting, no agent can directly observe the underlying state $s$, but each agent $i$ receives a private observation  $o^i \in \mathbb{O}_i$ correlated with the state. $\mathbb{O}$ is the set of joint observations, where each $\mathbb{O}_i$ is a set of observations available to agent $i$.  $O(o | a, s')$ is the probability of the joint observation $o$, given that the joint action $a$ led to state $s'$.  Agent reward $r_i: S \times A_1 \cdot \cdot \cdot A_N \rightarrow \mathbb{R}$ is a function of the state and joint action taken.  The objective is to infer a set of agent policies (the joint policy) that maximize the expected long-term \textit{shared} return $R$.    
\nocite{oliehoek2016concise}   

\subsection{Zero-Shot Coordination and Other-Play}
Zero-shot coordination (\textbf{ZSC}) is the problem setting of agents coordinating at test time with other agents who have been independently trained \citep{hu2020other}.  In cooperative multi-agent settings, a key challenge is to learn \textit{general} skills for coordinating and communicating with other agents.  Nonetheless, just as in single-agent settings, where agents can overfit to their environment, in multi-agent settings, agents can co-adapt with and overfit to their training partners. 
The goal of ZSC is to prevent this from happening, forcing the agents to instead learn policies that are compatible with novel partners (\textit{e.g.} new human or machine).

In order to enable generalization to independently trained partners \citep{hu2020other} introduce equivalence mappings $\Phi$, which characterize symmetries for each element in the Dec-POMDP. Defining: $\phi = \{\phi(S), \phi(A), \phi(O)\}$, where $\phi \in \Phi$, the equivalence implies the following:  
\begin{equation} \label{eq:phi_elements}
    \begin{split}
        \phi \in \Phi & \iff P(\phi(s') \mid \phi(s),\phi(a)) = P(s' \mid s,a) \\
        & \land R(\phi(s'),\phi(a),\phi(s)) = R(s',s,a) \\
        & \land O(\phi(o) \mid \phi(s),\phi(a),i) = O(o \mid s,a,i) 
    \end{split} 
\end{equation}
Since the notation is heavily overloaded, we can write shorthand as $\phi = \{\phi_S, \phi_A, \phi_O \}$.  
The equivalence mappings also act on trajectories and policies as follows:
\begin{equation} \label{eq:phi_policy}
    \begin{split}
        \phi(\tau_t^i) & = \{\phi(o_0^i), \phi(a_0^i), \phi(r_0^i), ..., \phi(o_t^i), \phi(a_t^i), \phi(r_t^i)\} \\
        \phi(\pi) & \iff \pi'(\phi(a) \mid \phi(\tau)) = \pi(a \mid \tau), \forall \tau,a 
    \end{split} 
\end{equation}
Taken together, the equivalence mappings imply that permuting actions and states with respect to their symmetries yields equivalent trajectories and thus does not change the return.  
Since agents trained independently have \textit{no way} of coordinating a priori to break symmetries consistently, \citep{hu2020other} introduce the \emph{other-play} (\textbf{OP}) objective:
\begin{equation} \label{eq:OP_objective}
    \pi^*_{OP} = \argmax_\pi \; \mathbb{E}_{\phi \sim \Phi} \;  J(\pi^1, \phi(\pi^2))
\end{equation}
Rather than maximizing for a specific joint policy, OP maximizes for the expected return over an \emph{equivalence class} of policies, when the policies for different players are sampled i.i.d. from this class.


\subsection{Referential Games}
Referential games have been extensively explored in the emergent communication literature \citep{lazaridou2016multi,havrylov2017emergence,lazaridou2018emergence,lowe2019pitfalls}; they require receivers to predict what senders are \textit{referring to}, through messages sent.  There are two learnable models: a sender network (for generating messages) and a receiver network (for observing and interpreting messages sent).  Intents are given as input to senders, but are not observable to receivers.  Messages can be composed of a single action or an entire trajectory (state-action sequence).  Sender generated messages are given as input to receivers.  The receiver outputs a distribution over the set of candidate intents.  Let cost $c \in \mathbb{R}$ represent cost of an action $a \in A$.  Referential games generally assume a cheap-talk setting, such that actions have no cost: $\{c(a) = 0 \; | \; \forall{a \in A}\}$.  They are also \textit{fully cooperative}, so senders and receivers are updated based upon a common payoff, the shared return. 

In this work, we employ a centralized mutli-agent training regime with decentralized execution \citep{foerster2016learning,lanctot2017unified,rashid2018qmix}.  Agents can share arbitrary information during training, \emph{e.g.} through exchanging gradients.  At test time however, agents are restricted to acting on their local action-observation history.

\subsection{Zipf Distribution}
The Zipfian distribution \citep{zipf2016human} was originally formulated in Quantitative Linguistics to characterize frequency (likelihood) of words, from natural language corpus data  \citep{piantadosi2014zipf}.  It is a discrete power law probability distribution describing the inverse correlation between rank and frequency; lower ranked items occur with higher frequency.  In this work, a Zipfian distribution is used to induce a nonuniform, monotonic distribution over the set of communicative intents, given as \textit{a priori} knowledge to agents.

\fi
	

\section{The Zero-Shot Communication Setting}
\label{sec:approach}

\iftrue
In this section, we describe our multi-agent problem setting for learning communication protocols.  The section contains three key ideas: (1) characterizes how a costly communication channel can be mapped to action equivalence classes exploitable for coordination, (2) establishes the utility of the OP objective for learning those equivalence classes, and (3) links the use of equivalence classes for communication protocol learning to information theory.

We first introduce the \textit{ZS Communication} (\textbf{ZSComm}) problem setting, as a specific instance of the Zero-Shot Coordination (\textbf{ZSC}) setting, where \textit{the coordination task for the agents is to effectively communicate intent with independently trained partners}.  
We use \textit{cross-play} as our evaluation metric, as is standard in the ZSC setting. Cross-play (\textbf{XP}) can be thought of as instance of out-of-distribution testing, since senders are paired with receivers who have been trained independently. 

In contrast to the standard cheap talk setting for referential games, our problem setting is inspired by embodied agents who \textit{implicitly} communicate through \textit{actions} taken. It assumes common-knowledge communication costs \textit{grounded} in the environment in which the agents exist: $\{c(a^i) \geq 0 \; | \; \forall{a^i \in A_i, i=1...N}\}$, where $c(a^i)$ is the cost of action taken by agent $i$.  As illustrated in Figure \ref{fig:embodied_referential_game}, costs are assigned to actions a priori, but agents are never directly given the costs, nor are ever made aware that any predefined, exploitable structure exists. They must \textit{discover} the cost-based latent structure underlying the communication protocol, through repeated play of the cooperative game.  
For the problem of learning communication protocols, we define a set of goals (intents) $G$ to be communicated and costs on agent actions $C(A)$. 


\subsection{Equivalences in Zero-Shot Communication}

As a baseline, protocol learning using self-play (\textbf{SP}) employs a learning objective $J$ that seeks to maximize utility of the joint action for \textit{training} partners. Let us consider a two-agent case where $\pi^1$ and $\pi^2$ represent individual agent policies and $\pi$ denotes the joint policy. SP maximizes the objective: 
\begin{equation} \label{eq:SP_objective}
    \pi^*_{SP} = \argmax_\pi J(\pi^1, \pi^2) = \argmax_\pi J(\pi^2, \pi^1)
\end{equation}
For referential communication, agent 1 receives as input a goal $g \sim G$ to be communicated, as part of its private observation.  It selects an action $a^1 \in A_1$ as its message.  Agent 2 observes the message and must predict the goal as its action.  Thus the SP objective can be specified for the referential task as:
\|*
\begin{align}
        \pi^*_{C-SP} &= \argmax_\pi J(\pi^2(G \mid A_1, \mathbb{O}_2), \pi^1(A_1 \mid G, \mathbb{O}_1)) \label{eq:SP_obj_pt1}
\end{align}
*/
\begin{align}
        \pi^*_{C-SP} &= \argmax_\pi \; \mathbb{E}_{g \sim G, o \sim \mathbb{O}, a \sim \pi} \; J(\pi^2(g \mid a, o^2), \pi^1(a \mid o^1, g)) \label{eq:SP_obj_pt1}
\end{align}
Given Equation \ref{eq:SP_obj_pt1} maximizes the joint action taken by sender and receiver, this implies inferring protocols with a maximally informative relationship between actions and goals, of the \textit{training} pair. 

One challenge though is when there are several actions in the same equivalence class, the way the action symmetries are broken (\textit{i.e.} selection between equivalent actions is decided) is then specific to the joint policy of the \textit{training} agents.  Nonetheless, we seek to enable agents to learn communication policies that generalize to independently trained agents (\textit{novel} partners) at test time.  Thus, agents should also be prepared to partner at test time with symmetry-equivalent policies. 

Thus, using the OP objective (Equation \ref{eq:OP_objective}), we can infer protocols that generalize \textit{beyond} training partners by replacing sender policy $\pi^1$ in Equation \ref{eq:SP_obj_pt1} with $\phi(\pi^1)$: 
\|*
\begin{equation} \label{eq:OP_emecomm_objective}
    \pi^*_{C-OP} = \argmax_\pi \; \mathbb{E}_{\phi \sim \Phi} \; J(\pi^2(G \mid A,\mathbb{O}_2), \pi^1(\phi(A_1) \mid G, \phi(\mathbb{O}_1))
\end{equation}
*/
\begin{equation} \label{eq:OP_emecomm_objective}
    \pi^*_{C-OP} = \argmax_\pi \; \mathbb{E}_{g \sim G, \phi \sim \Phi, o \sim \mathbb{O}, a \sim \pi} \; J(\pi^2(g \mid a, o^2), \pi^1(\phi_a(a) \mid \phi_o(o^1), g))
\end{equation}
For agent 2, this removes dependence upon a specific agent 1 policy (used for training) and instead allows inference of communicative goal from an entire equivalence class of agent 1 policies.  


From an information theoretic perspective, the OP objective induces a relationship $I(G;A) = I(G;\Phi(A)) \gg 0$, yielding learned protocols with high mutual information between $G$ (output of $\pi^2$) and $\Phi(A)$ (output of $\pi^1$).  
This implication is important for ZSComm; it suggests recovering the \textit{equivalence class} of an action is sufficient to infer the communicative \textit{goal}.   
To exploit equivalence mappings for learning ZSComm protocols however, we must first \textit{discover} them. 

\fi


\section{Discovering Equivalences}
\label{sec:approach}

\iftrue

\vspace{-1mm}
To learn  unique, optimal ZSC policies, our algorithm must automatically \textit{discover} equivalence mappings (symmetries) $\Phi$ for the underlying Dec-POMDP.  
In general, discovering symmetries in a Dec-POMDP is an NP-hard problem since they correspond to graph-automorphisms~\citep{lubiw1981some}. Our key insight is that for the purpose of ZSC we do not need to discover all symmetries. Instead, we only need to discover those that lead to \emph{equivalent but mutually incompatible} optimal joint policies. 

Furthermore, rather than having to search over symmetries in the Dec-POMDP, we can directly discover them from these \emph{optimal policies}. If $\pi^*$ is an optimal policy, by definition, $\phi(\pi^*)$ is also an optimal policy for any symmetry mapping $\phi$. Thus, a learning process that produces $\pi^*$ in training run, will produce all equivalent policies $\phi(\pi^*)$ with the same probability in future runs. 

Based on these insights, we introduce \textit{Quasi-Equivalence Discovery} (\textbf{\ourmethod{}}), an iterative algorithm for discovering symmetries and exploiting them to infer optimal ZSC protocols.  At a high level, for each iteration the algorithm runs, there are two steps involved in optimizing agent policies: (1) use the current set of equivalence mappings as input to train a new set of optimal joint policies under the ZSC setting, and (2) recover \textit{all} equivalence mappings from the trained optimal policies.  Convergence to a \textit{unique} ZSC policy occurs when $SP == XP$ for the set of independently trained joint policies.  The remainder of this section describes the algorithm in more depth.  Algorithm~\ref{alg:ourmethod} details \ourmethod{}.


\setlength{\intextsep}{10pt}
\begin{algorithm}[th]
    \begin{algorithmic}[1]
    \footnotesize{
        \STATE{Let $\Pi \gets$ set of N joint policies $\{ \pi_1,... \pi_N \}$ }
        \STATE{Initialize symmetries to identity $\Phi^* = \{ \mathcal{I}\} \;$  (as per line 228)}
        \WHILE{$XP(\Pi^{*}) < SP(\Pi^{*}) - \epsilon$}
        \STATE{Train new set of $N$ \textit{optimal} joint policies $\Pi^*$ equivariant under $\Phi^*$}
        \STATE{Initialize set of candidate equivalence mappings randomly: $\Phi=\{\phi^{i,j}\},$ $i,j=1... N$, $\phi^{i,j} =\{\phi^{i,j}_o, \phi^{i,j}_a\}$ }
        \STATE{$\#$ Iterate through \textit{all pairs} of optimal joint policies and extract symmetries}
        \FOR{$\pi_i \in \Pi^*$}
            \FOR{$\pi_j \in \Pi^*  \mid  j \neq i$}
                \STATE{Update $\phi^{i,j}$ to minimize KL Divergence between $\pi_i $ and $\pi_j$  (where $\phi(\tau)$ defined in Equation \ref{eq:phi_policy}):
                \begin{align}
                    L(\phi^{i,j}) &= \mathbf{E}_{\tau \sim P(\tau \mid \pi_i)} \; D_{KL}\left(\pi_i(a \mid \tau) \parallel \pi_j( \phi^{i,j}_a(a) \mid \phi^{i,j}(\tau) \right)\\   
                    &= \mathbf{E}_{\tau \sim P(\tau \mid \pi_i)} \; \sum_a \pi_i^k(a \mid \tau^k) \log \bigg(\frac{\pi_i^k(a \mid \tau^k)}{ \pi_j^k( \phi^{i,j}_a(a) \mid \phi^{i,j}(\tau^k)} \bigg); \; k= 1,2
                \end{align}}
                \STATE{If $L(\phi^{i,j}) < \delta$: Append $\phi^{i,j}$ to $\Phi^*$ }
            \ENDFOR
        \ENDFOR
        \ENDWHILE
        \STATE{\textbf{return} $\Pi^*[0]$}
    }
    \end{algorithmic}
    \caption{\ourmethod~Algorithm}
    \label{alg:ourmethod}
\end{algorithm}
Equivalence mappings (EM) in the QED algorithm operate on observations, actions, and trajectories, where trajectories are (observation, action) sequences. EMs are transformation matrices that map one entity of a set to another (equivalent) entity of the same set.  For example, action EMs $(\phi_a)$ transform one action $(a_j)$ into another $(a_i)$ by permuting the first: $a_i \gets \phi^{i,j}_{a}(a_j) := \phi^{i,j}_{a} \times a_j$.  
The same is true for observation EMs $(\phi_o)$.  For trajectories, EMs are employed for each individual observation and action in the sequence, like in Equation \ref{eq:phi_policy}.

The set of EMs being learned, $\Phi^*$, is initialized with square \textit{Identity} matrices, of sizes $|\mathbb{O}|$ and $|A|$, for observations and actions respectively.  
Each iteration proceeds as follows.  
First, \ourmethod{} trains a new set of optimal joint policies that are equivariant under the current set of EMs $\Phi^*$.
For ZSComm tasks, we use an OP objective for this step (Equation \ref{eq:OP_emecomm_objective}).  
Second, \ourmethod{} updates the set of equivalence mappings using the learned \textit{optimal} policies from the previous step.  It does this by iterating through each pair of joint policies $\pi^{*}_i, \pi^{*}_j \in \Pi^{*}$ and inferring a mapping $\phi^{i,j}$ that transforms $\phi^{i,j}(\pi^{*}_j)$ to $\pi^{*}_i$.  The learned mapping is added to the set of mappings $\Phi^*$, once loss for it is sufficiently low.  
Finally, \ourmethod{} uses the convergence to a unique optimal ZSC joint policy as the stopping criterion.  It completes when SP performance approximately equals XP performance, for the set $\Pi^*$.  
\fi


\section{Experiments and Results}
\label{sec:experiments}

\iftrue
\vspace{-2mm}

\begin{table*}[t]
 	\centering
 	\begin{subtable}[b]{\textwidth}
 	    \centering
        \begin{tabular}{ |p{3cm}||p{2cm}|p{2cm}||p{2cm}|p{2cm}| }
         \hline
         &\multicolumn{2}{|c||}{\textit{No} Degeneracy Task}& \multicolumn{2}{|c|}{Energy Degeneracy Task} \\ \hline
         &Training (SP) &Test (XP) &Training (SP) &Test (XP)\\ \hline \hline
         Max Class (Baseline) & 0.44 & 0.44 & 0.44 & 0.44\\ \hline
         SP & 0.88 $\pm$ 0.089 & 0.42 $\pm$ 0.293 & \textbf{0.95} $\pm$ 0.007 & 0.24 $\pm$ 0.217 \\ \hline  
         SP w/ max filtering & \textbf{0.96} $\pm$ 0.004 & \textbf{0.96} $\pm$ 0.009 & \textbf{0.96} $\pm$ 0.002 & 0.57 $\pm$ 0.112\\ \hline
         OP & \textbf{0.96} $\pm$ 0.001 & \textbf{0.96} $\pm$ 0.001 & 0.93 $\pm$ 0.003 & \textbf{0.93} $\pm$ 0.003\\ \hline
         \ourmethod{} & \textbf{0.96} $\pm$ 0.001 & \textbf{0.96} $\pm$ 0.001 & 0.93 $\pm$ 0.003 & \textbf{0.93} $\pm$ 0.003\\ \hline
        \end{tabular}
    	\label{tab:zs_coord_2_no_curr}
	\end{subtable}
 	\caption{\textbf{Performance on Communication Tasks.} Comparison of methods for training optimal ZSComm policies. SP, OP given the ground truth symmetries, and \ourmethod{} are all expected to perform comparably well for the \textit{No} Degeneracy task, because there is one \textit{unique} optimal SP policy, using any of these training algorithms.  The \textit{No} Degeneracy task (left) is intended to illustrate that even when SP is at its best, OP given the symmetries, and \ourmethod{} can match its performance.  For the Energy Degeneracy task, there are \textit{many} optimal SP policies, but one \textit{unique} optimal policy, using OP (given the symmetries) or \ourmethod{}. The Energy Degeneracy task (right) is intended to illustrate that \ourmethod{} will \textit{both} discover the correct symmetries \textit{and} use them to significantly outperform SP for ZSComm, in situations where there is redundancy in the action space (\textit{e.g.} many trajectories that exert same amount of energy), and agents have not been allowed to pre-coordinate on how to break action symmetries.}   
 	\label{tab:overall_task_performance}
\end{table*}

\begin{table*}[t]
 	\centering
 	\begin{subtable}[b]{\textwidth}
 	    \centering
        \begin{tabular}{ |p{3.5cm}||p{4.0cm}|p{4.0cm}| }
         \hline
         \multicolumn{3}{|c|}{\textit{Ablation Experiments:} Energy Degeneracy Task (Test: XP)} \\ \hline
         &\textit{Uniform} Goal Distribution &\textit{Zipfian} Goal Distribution\\ \hline \hline
         Max Class (Baseline) & 0.20 & 0.44\\ \hline
         Cheap Talk Channel & 0.20 $\pm$ 6.4e-5 & 0.27 $\pm$ 9.5e-5 \\ \hline  
         \textit{Costly} Channel & 0.20 $\pm$ 6.3e-7 & \textbf{0.93} $\pm$ 0.003\\ \hline
        \end{tabular}
	\end{subtable}
 	\caption{\textbf{Ablation Experiments for \ourmethod{} Algorithm.}  \textit{Energy Degeneracy} Task: (Cheap Talk, Costly) $\times$ (Uniform, Zipfian). Comparison of four ablations of \ourmethod{}, removing each common knowledge constraint and analyzing how this change in problem setting impacts performance. This strong degradation in performance (for the three ablations) happens because all actions are placed in the same equivalence class and thus all actions are interchangeable.  So the optimal strategy is to predict a highest likelihood goal at \textit{random} (top row).  Since however performance reflects the joint \textit{confidence}, if probability mass is distributed across different actions for any given goal, this will take some mass away from the most likely goal.  \textit{Overall:} Illustrates that \textit{both} common knowledge constraints (bottom right) are \textit{critical} for inferring an optimal ZSComm policy.}   
 	\label{tab:ablation_experiments}
 	\vspace{-4mm}
\end{table*}

\begin{figure*}[tb]
	\centering
	\begin{adjustbox}{minipage=\linewidth,scale=0.95}
		\centering
		\begin{subfigure}[b]{0.48\textwidth}
			\centering
			\includegraphics[width=\textwidth]{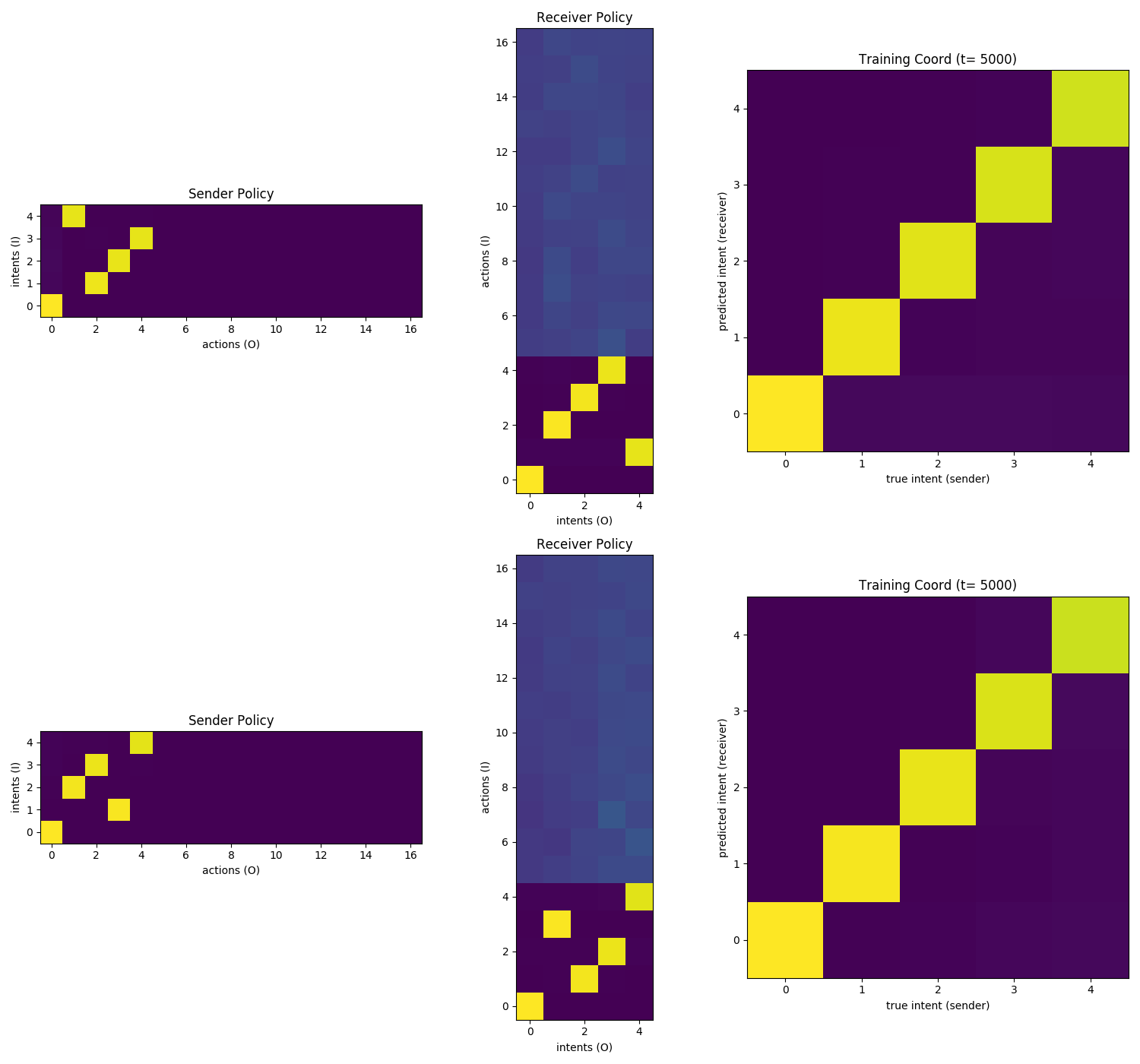}
			\caption{SP: Training Pairs Performance (SP)}
			\label{fig:sp_task2_sp}
		\end{subfigure}
		\begin{subfigure}[b]{0.48\textwidth}
			\centering
			\includegraphics[width=\textwidth]{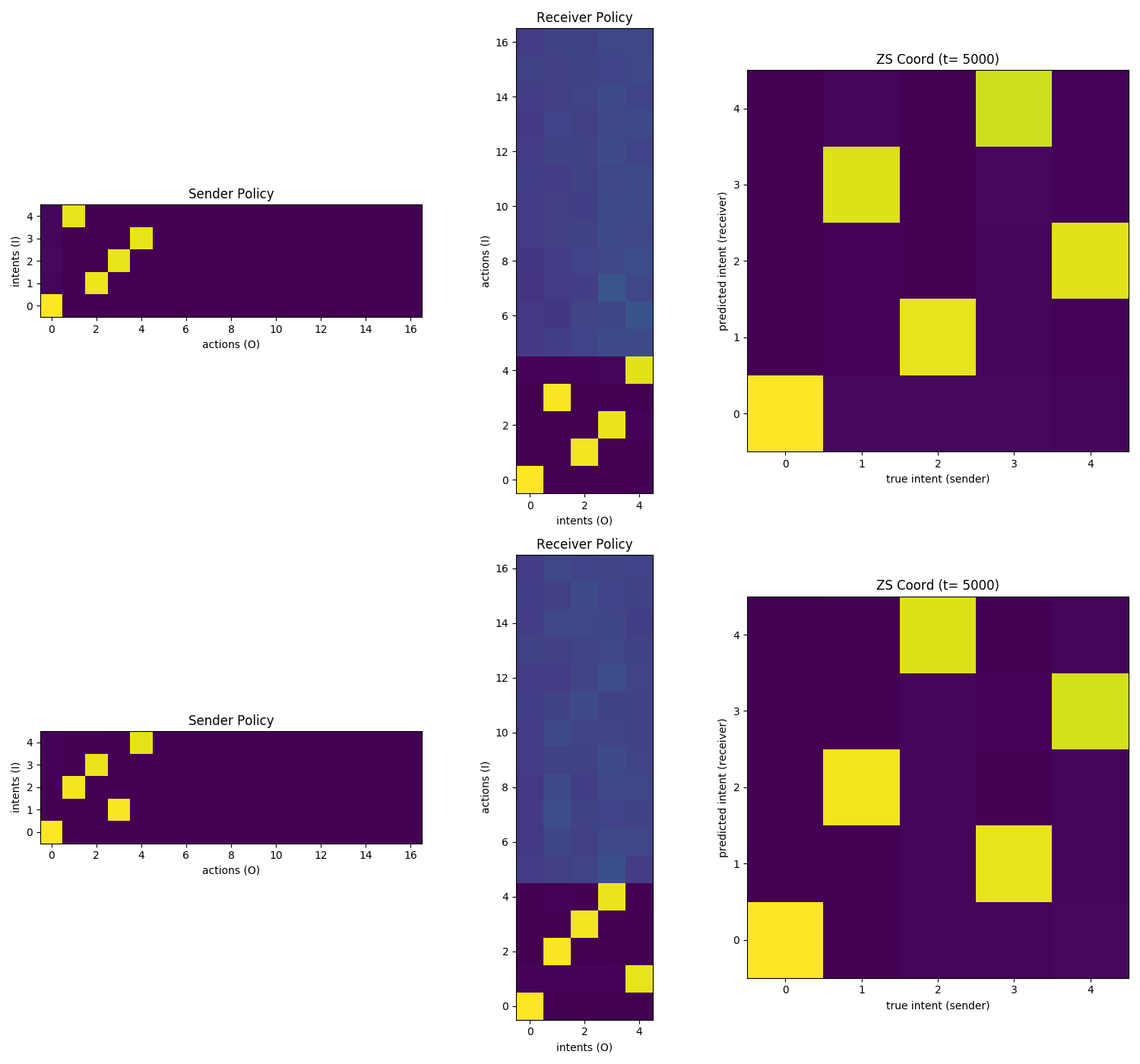}
			\caption{SP: \textit{Novel} Pairings Tested (XP)}
			\label{fig:sp_task2_xp}
		\end{subfigure}
	\end{adjustbox} 
	\caption{\textbf{SP Policies} Learned. \textit{Energy Degeneracy} Task. Shows performance for two agent pairs (top, bottom) -- with Training Partners (left) and Evaluated for ZSC (right). Left column shows sender policy mapping given goals to selected actions. Middle column shows receiver policy mapping sender actions to predicted goals. Right column shows SP Performance ($p(predictedGoal \; | \; trueGoal)$) at the end of Protocol Training. Coordination success is depicted by probability mass along the diagonal of the confusion matrix (third column, SP and XP). Illustrates that using optimal SP polices achieves \textit{almost perfect} performance w/training partner (left), as shown by the XP confusion matrix diagonal, \textit{but} performance degrades significantly for ZSC (right).}
	\label{fig:task2_SPpolicies_spxp} 
    \vspace{-3mm}
\end{figure*}

\begin{figure*}[tb]
	\centering
	\begin{adjustbox}{minipage=\linewidth,scale=0.95}
		\centering
		\begin{subfigure}[b]{0.48\textwidth}
			\centering
			\includegraphics[width=\textwidth]{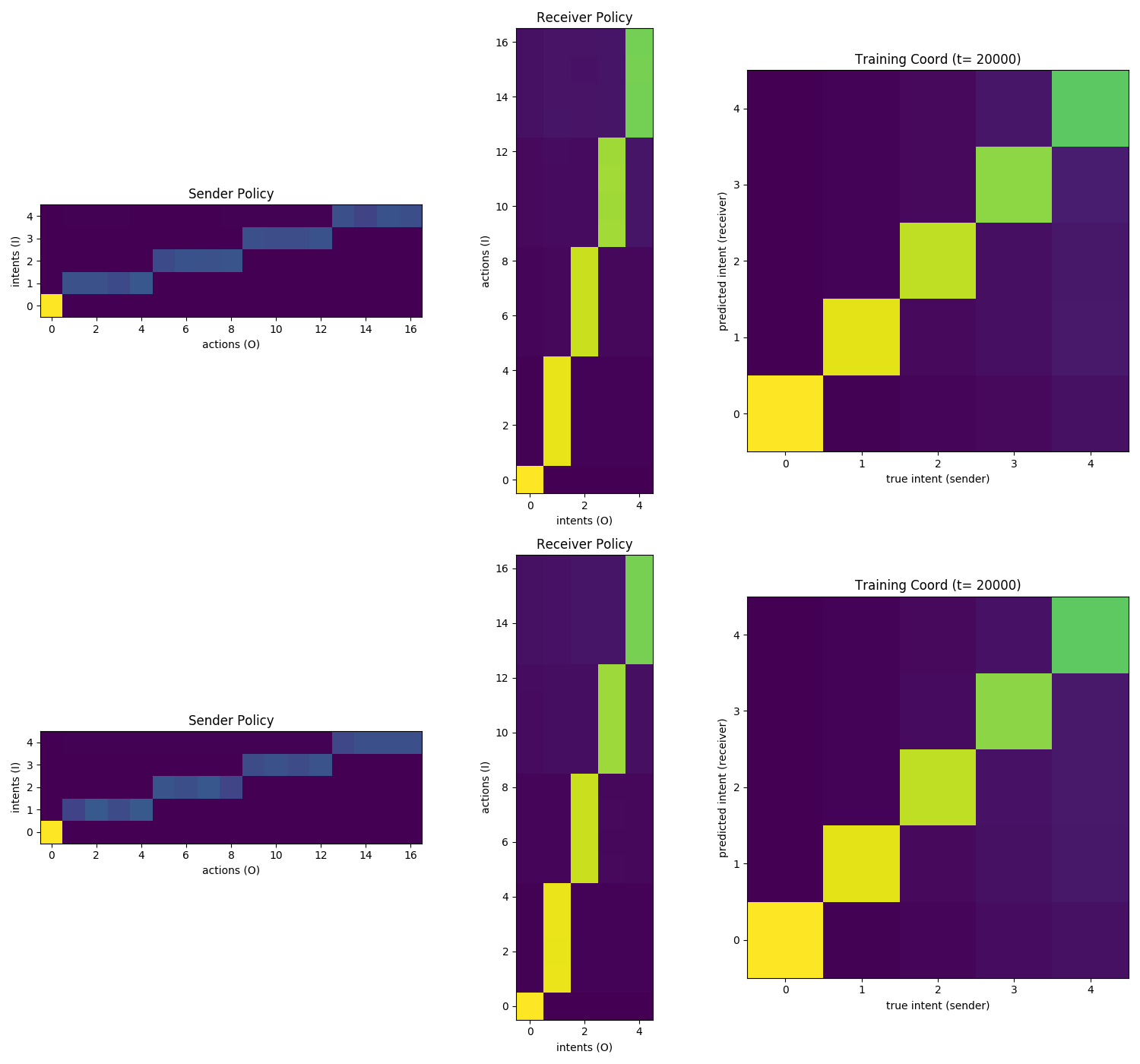}
			\caption{\ourmethod{}: Training Pairs Performance (SP)}
			\label{fig:op_task2_sp}
		\end{subfigure}
		\begin{subfigure}[b]{0.48\textwidth}
			\centering
			\includegraphics[width=\textwidth]{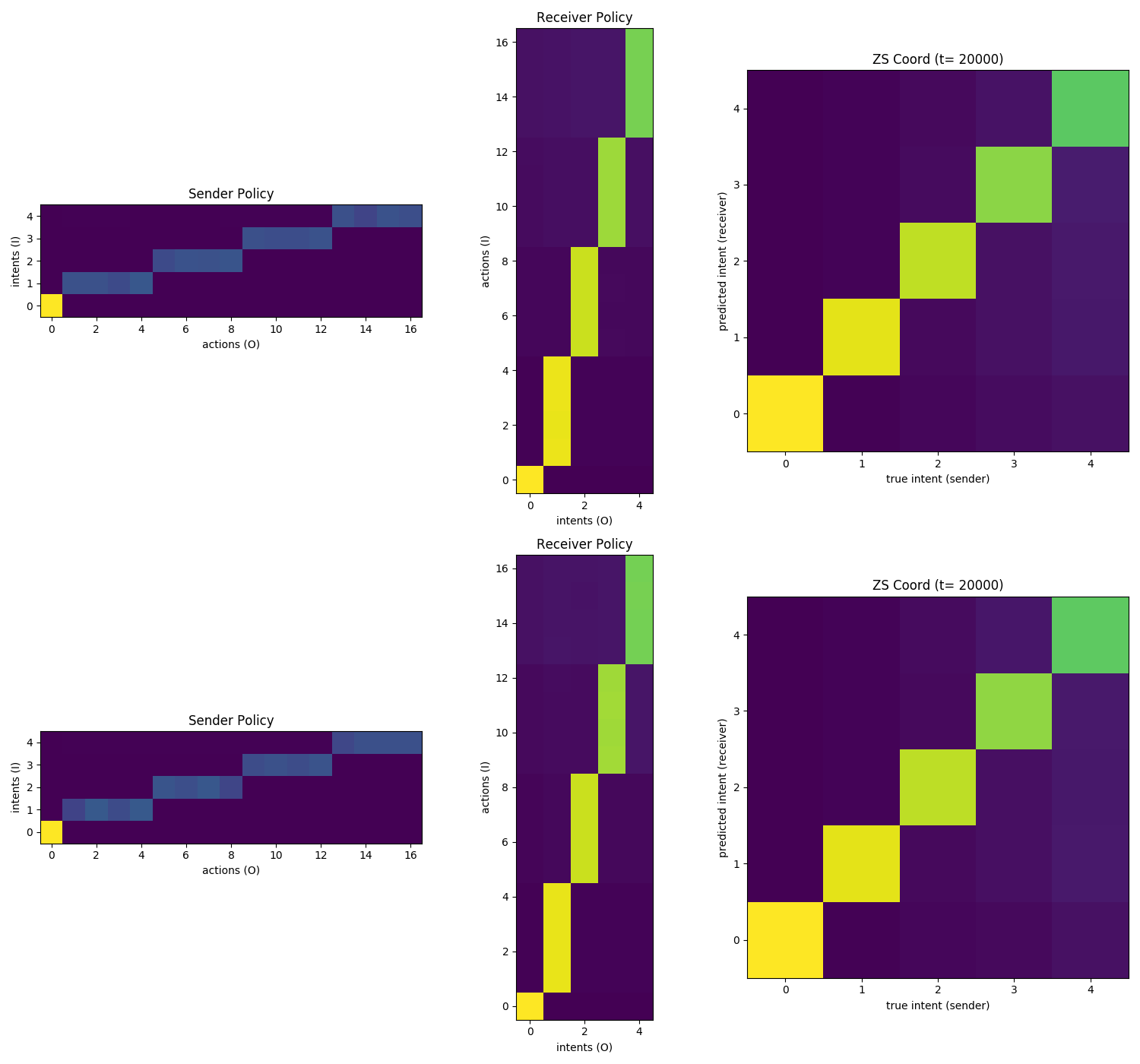}
			\caption{\ourmethod{}: \textit{Novel} Pairings Tested (XP)}
			\label{fig:op_task2_xp}
		\end{subfigure}
	\end{adjustbox} 
	\caption{\textbf{\ourmethod{} Policies} Learned. \textit{Energy Degeneracy} Task. Same figure description as above.  Illustrates that \textit{consistently high} performance, \textit{both} in SP and XP, derives from there being a \textit{unique} optimal policy. So even when paired with a novel partner in XP, the paired optimal policies are compatible, as shown by the XP diagonal.}  
	\label{fig:task2_OPpolicies_spxp} 
    \vspace{-5mm}
\end{figure*}

\subsection{Experimental Setup}  \label{sec:experimental_setup}
\vspace{-2mm}

In this section, we describe the communication tasks used for comparing performance of \ourmethod{} with SP and OP given the exact symmetries.  All methods use the same combination of accuracy and energy cost for training policies according to their specific learning objective.  We evaluate on two tasks, both with 5 goals, but of increasing complexity in the communication action space.  
We also contribute a colab notebook as an instructive tool for exploring protocol learning in our simple embodied domain\footnote{Colab notebook for \textit{Costly}-Channel Communication Protocols -- http://shorturl.at/luHPX}. 

\textbf{\textit{No} Degeneracy Task.} This task considers 10 discrete actions, \textit{each} assigned its own unique energy value: $C(a) = id(a)$.  For achieving \textit{both} maximal communication efficacy \textit{and} minimal cost, each goal must be mapped to the lowest energy value it can be assigned, but no two goals can be mapped to the same action.  In particular, to produce a minimal effort protocol, the frequently occurring goals be mapped to lowest cost actions.  There is only \textit{one} globally optimal protocol for this task.

\textbf{Energy Degeneracy Task.} This task also consider protocols with five distinct energy values, but with 17 discrete actions, i.e., with action redundancy.  In particular, one action has a unique, lowest cost; the other 16 actions are partitioned into four sets of 4, where each \textit{set} is mapped to a distinct cost.  
This experiment investigates what happens in the degenerate case where several actions can be mapped to the same energy cost.  Thus even in optimizing for minimal effort protocols, there are \textit{multiple} ways to converge upon a globally optimal protocol, and consequently, ZSC is significantly more challenging.  
This redundancy creates a need to break action symmetries.  This type of communication complexity gets closer to what may be expected (and exacerbated even further) in continuous, embodied domains.

\textbf{Representation.}  The sender and receiver policies have different state, observation, and action spaces.  
The sender policy takes as input a batch of $|G|$ one-hot vectors, each of size $|G|$.  Its local observations are thus only ground-truth communicative goals.  Each goal is weighted by its likelihood under a Zipfian distribution.  Each goal is uniquely identified by its rank in the Zipfian.  The sender's action space is the set of communicative actions $A$.  
The receiver policy takes as input a batch of $|A|$ one-hot vectors, each of size $|A|$.  Its local observations are composed of sender actions.  The receiver's action space is the set of goals $G$.  It can make predictions based on observations of sender action.

\textbf{Training and Evaluation.} Agent policies are single-layer perceptrons: feed-forward neural networks with no hidden layer and a softmax function on the output layer.  The sender policy outputs a distribution over actions for each ground truth goal; the receiver policy outputs a distribution over goals for each communication action.   
For training agent policies, we compute the exact policy gradient using the Cross-Entropy between the \textit{ground-truth} goal being communicated by Agent 1 and the \textit{predicted} goal by Agent 2: $R = \mathbb{E}_{g \sim G} \; -\log \; p_\pi (\hat{g} = g \mid g)$.  An entropy term was used to encourage exploration.  Energy cost was computed as: $\mathbb{E}_{g \sim G} \; C(a) \; p_{\pi^1}(a \mid g)$. 
For all experiments, the weights used for the entropy loss and the energy cost were $1e-2$ and $3e-1$, respectively.  We used stochastic gradient descent for optimization, with a learning rate of $0.1$.  The shared return is backpropagated from Agent 2 to Agent 1.  On the \textit{no-degeneracy} task, training algorithms each ran for 5000 iterations. On the \textit{energy degeneracy} task, they each ran for 10000 iterations.  We filter the maximum $k\%$ of policies to obtain \textit{only} near optimal policies.  Through empirical tuning, we set $k=10$. Evaluation of communication efficacy with training partners is validated with SP, and ZSComm is evaluated as XP performance.  

\textbf{Ablations.} We perform ablation experiments to analyze the effect of each common knowledge constraint assumed in our problem setting.  The ablations examine how communication efficacy using \ourmethod{} changes on the (more complex) Energy Degeneracy task as constraints are removed. The conditions are a cross product of: (Cheap Talk Channel, Costly Channel) $\times$ (Uniform Distribution, Zipfian Distribution).  The former relaxes the assumption of a costly channel, where cheap talk is the baseline; the latter perturbs the distribution over goals, where uniform is the baseline.

Additional experimental details and compute resources used are discussed in the Appendix.

\subsection{Results}
\vspace{-2mm}

\textbf{Communication Task Overall Performance.} Our first experiment compares communication efficacy of \textit{all} methods investigated (SP, OP, \ourmethod{}) on both communication tasks: No Degeneracy and Energy Degeneracy. Table \ref{tab:overall_task_performance} lists mean and standard standard across ten agent pairs.  
This experiment contributes our main result.  The \textit{No} Degeneracy task shows that even when SP with filtering (selecting the maximum $k$ policies, where $k$ is a hyper-parameter) is sufficient to learn an optimal ZSComm policy, \ourmethod{} can perform comparably well and matches the performance of OP with the ground truth symmetries \textit{given}. The Energy Degeneracy task provides evidence that SP, even with filtering, is not sufficient to achieve ZSComm on more complex communication tasks.  Here, although SP has the highest training performance (though by a marginal amount), it is substantially outperformed by \ourmethod{} once redundancy in the action space is present.  Again, \ourmethod{} is able to match the performance of OP given the ground truth action symmetries.

\textbf{Proposed Algorithm (\ourmethod{}) on Ablation Experiments.} Our second experiment seeks to understand the effect of each of the common knowledge constraints assumed in our problem setting: (1) energy costs and (2) a nonuniform distribution over goals.  Table \ref{tab:ablation_experiments} shows results from four ablations, relaxing each of these assumptions.  The key insight from this experiment is that \textit{both} constraints are critical for learning ZSComm policies.

\textbf{Qualitative Analysis on Energy Degeneracy Task.}  Finally, we visually inspect agent policies for the more complex degeneracy task.  This is to better understand why SP performance degrades as communication complexity increases and how \ourmethod{} was able to so significantly outperform SP, once redundancy in the action space was present.  Figures \ref{fig:task2_SPpolicies_spxp} and \ref{fig:task2_OPpolicies_spxp} show policies learned for SP and \ourmethod{}, respectively.  
The key insight is that there are multiple optimal SP policies in the Energy Degeneracy task, because there are many ways to permute equivalent actions and result in an optimal SP policy.  With no way to pre-coordinate on how to break action symmetries however, ZSComm is extremely challenging.  \ourmethod{} converges on a unique OP policy. Thus even with an independently trained partner at test time, the polices are compatible.

Overall, our experimental findings provide strong evidence that \ourmethod{} is able to leverage the OP learning objective to both: (1) \textit{discover} equivalence mappings and (2) \textit{emerge} more \textit{general} communication protocols, capable of coordinating with \textit{novel} partners.  

\fi

\section{Conclusions and Future Work}
\label{sec:conclusion}

\iftrue
\vspace{-2mm}
We introduce a novel emergent communication setting that combines costly messages with non-uniform priors over intents. Crucially, in our setting, we can meaningfully study \emph{zero-shot emergent communication}, since the common-knowledge problem setting leads to unique optimal ZSC policies. However, we also show that current self-play methods fail to learn these policies and fail to coordinate at test-time. 
Next we introduced \ourmethod~which iteratively discovers the equivalence classes of policies in this setting and can indeed learn the optimal ZSC policies. 

Our work opens the door to future investigations along three main axes: First of all, this is the first instance of a successful ZSC method that can directly discover symmetries from interacting with the environment. This clearly is a hugely important direction and needs to be generalized to other environments and tasks.
Secondly, our costly communication game is an \emph{abstract proxy} for embodied agents, since they have to physically move their bodies and incur an energy penalty. However, scaling to high-dimensional, continuous action spaces and states is clearly important in order for this line of work to become applicable to realistic settings. 
Lastly, our work highlights the challenges of learning communication protocols and illustrates the crucial problems that sub-optimal protocols can cause for ZSC: While it is always undesirable to learn sub-optimal polices, in self-play settings the amount of damage done is directly related to the level of sub-optimality. So as long as solutions with high loss can be avoided, the test-time performance cannot be arbitrarily bad. In contrast, in ZSC, sub-optimal solutions can lead to complete \emph{coordination failure}, or even worse, for \ourmethod~to the learning of wrong equivalence classes.   
As such, finding methods that can learn \emph{optimal} communication protocols is of crucial importance for future work.

\fi



\nocite{van2011social,zipf2016human}

\bibliography{multiagent_comm}
\bibliographystyle{plainnat}

\clearpage

\appendix
    \|*
\section{Do \emph{not} have an appendix here}

\textbf{\emph{Do not put content after the references.}}
Put anything that you might normally include after the references in a separate
supplementary file.

We recommend that you build supplementary material in a separate document.
If you must create one PDF and cut it up, please be careful to use a tool that
doesn't alter the margins, and that doesn't aggressively rewrite the PDF file.
pdftk usually works fine. 

\textbf{Please do not use Apple's preview to cut off supplementary material.} In
previous years it has altered margins, and created headaches at the camera-ready
stage.

\subsection{Experimental Task 1: \textit{No Degeneracy} -- Additional Results}
\label{sec:discrete_task1_contd}
\begin{itemize}
    \item Brief recap of simplest task: \textit{no energy redundancy} - (zipf only vs zipf + energy)
    \item Examples of learned sender/receiver policies visualized
    \item Summary of SP and XP performance [mean/std dev]
\end{itemize}

\subsection{Experimental Task 2: \textit{Energy Degeneracy} -- Additional Results}
\label{sec:discrete_task2}
\begin{itemize}
    \item Brief recap of more complex task: \textit{energy degeneracy} - (zipf only vs zipf + energy)
    \item Examples of learned sender/receiver policies visualized
    \item Summary of SP and XP performance [mean/std dev]
\end{itemize}

*/

\iftrue

\section{Appendix}

\subsection{Theoretical Analysis of Communication Objective Function}

Let us consider the two-agent case where $\pi^1$ and $\pi^2$ represent individual agent policies, and $\pi$ denotes the joint policy $\pi = \{\pi^1, \pi^2\}$.  

SP maximizes the following objective: 
\begin{equation} \label{eq:SP_objective}
    \pi^*_{SP} = \argmax_\pi J(\pi^1, \pi^2) = \argmax_\pi J(\pi^2, \pi^1)
\end{equation}

In contrast, OP adapts the SP objective to maximize over equivalence classes of policies: 
\begin{align} \label{eq:OP_objective}
    \pi^*_{OP} &= \argmax_\pi \; \mathbb{E}_{\phi \sim \Phi} \;  J(\pi^1, \phi(\pi^2)) \\
    &= \argmax_\pi \; \mathbb{E}_{\phi \sim \Phi} \;  J(\phi(\pi^1), \pi^2) \label{eq:OP_obj_pt2}
\end{align}

\|*
In the referential game, agent 1 receives as input a goal $g \sim G$ to be communicated as part of its private observation, it selects an action $a^1 \in A_1$ as its message, agent 2 observes the message and must predict the goal as its action.  Thus, Equation \ref{eq:SP_objective} can be specified for the signaling task as:
\begin{equation}
        \pi^*_{C-SP} = \argmax_\pi \; \mathbb{E}_{g \sim G, o \sim \mathbb{O}, a \sim \pi} \; J(\pi^1(a \mid o^1, g), \pi^2(g \mid a, o^2)) \label{eq:SP_obj_pt1}
\end{equation}

Using the OP objective though, we can infer protocols that generalize \textit{beyond} training partners by replacing sender policy $\pi^1$ in Equation \ref{eq:SP_obj_pt1} with $\phi(\pi^1)$: 
\begin{equation} \label{eq:OP_emecomm_objective}
    \pi^*_{C-OP} = \argmax_\pi \; \mathbb{E}_{g \sim G, \phi \sim \Phi, o \sim \mathbb{O}, a \sim \pi} \; J(\pi^2(g \mid a, o^2), \pi^1(\phi_a(a) \mid \phi_o(o^1), g))
\end{equation}
*/

For the \textbf{Referential} Task: The primary objective is for agent 2 (receiver) to \textit{predict} what communicative goal agent 1 (sender) is \textit{signaling} or \textit{referring to}.  Thus the joint policy $\pi$ is optimized to make accurate predictions $\forall g \in G$ given a set of goals sampled from $G$ as input to the multi-agent system.  For a costly-channel setting, it is also jointly optimized to minimize cost.

In the derivation below, we consider only a simplified setting, where messages $:=$ individual actions.  \\

Mathematical Derivation for OP Objective (Equation \ref{eq:OP_objective}) on \textbf{Costly-Channel Referential} Task:
\begin{align} \label{eq:derivation}   
    \mathbb{E}_{\phi \sim \Phi} \; J( \pi^1, \phi(\pi^2)) &= \mathbb{E}_{\phi \sim \Phi, g \sim G} \left[ \log p_{\pi}(g \mid g) - cost(\pi) \right] \\
    &= \mathbb{E}_{\phi \sim \Phi} \left[ \sum_{g \in G} p(g) \; \log p_{\pi}(g \mid g) - \sum_{g \in G} C(a) \; \pi^1(a \mid g) \; p(g) \right] \\
    &= \mathbb{E}_{\phi \sim \Phi} \left[ \sum_{g \in G} p(g) \; \log \Bigg[  \sum_{a \in \mathcal{A}} \pi^2(g \mid \phi(a)) \; \pi^1(a \mid g) \Bigg] - \sum_{g \in G} C(a) \; \pi^1(a \mid g) \; p(g) \right] \\
    &= \sum_{g, \phi} p(g, \phi) \; \log \left[ \sum_{a \in \mathcal{A}} \pi^2(g \mid \phi(a)) \; \pi^1(a \mid g) \right] - \sum_{g} C(a) \; \pi^1(a \mid g) \; p(g) \\
    &\geq \sum_{g, \phi, a} p(g, \phi) \; \log \left[\pi^2(g \mid \phi(a)) \; \pi^1(a \mid g) \right] - \sum_{g} C(a) \; \pi^1(a \mid g) \; p(g) \\
    &= \sum_{g, \phi, a} p(g, \phi) \; \log \pi^2(g \mid \phi(a)) + \sum_{g,a} p(g) \; \log \pi^1(a \mid g) - \sum_{g} C(a) \; \pi^1(a \mid g) \; p(g) \\
    &=\sum_{a, g, \phi} p(g, \phi) \; \log \pi^2(g \mid \phi(a)) + |\mathcal{A}| \; \mathbb{E}_{g \sim G} \sum_{a} p(a) \log \pi^1(a \mid g) - \sum_{g} C(a) \; \pi^1(a \mid g) \; p(g) \\
    &= -\sum_{a}  H(G \mid \Phi(a)) \;  - |\mathcal{A}|\; \mathbb{E}_{g \sim G} \left[H\bigg( p(a), \pi^1(a \mid g) \bigg) \right] - \mathbb{E}_{a \sim \pi^1, g \sim G} \; C(a) \label{eq:derivation_final_line} 
\end{align} \\  

For computing mutual information: $I(G;\Phi) := H(G) - H(G|\Phi)$, where $I$ represents \textit{mutual information} and $H$ represents \textit{entropy}.  However, the distribution over goals is given and stationary, so $H(G)$ is held constant.  Thus for inferring an \textit{optimal} joint policy $\pi^*$, Equation \ref{eq:derivation_final_line} implies:
\begin{align} \label{eq:derivation_continued}
    \pi^{*} &= \argmax_{\pi} \; \left[\sum_{a} I_{\pi}(G;\Phi(a)) \; - |\mathcal{A}|\; \mathbb{E}_{g \sim G} \left[H\bigg( p(a), \pi^1(a \mid g) \bigg) \right] \; - \mathbb{E}_{a \sim \pi, g \sim G} \; C(a) \right] 
\end{align}


\textit{Implications of Derivation.} 
Using the OP objective for a Costly-Channel Referential Task induces an \textbf{optimal} protocol with the following important properties: (1) maximizes mutual information between \textit{goals} and \textit{equivalence classes} over actions, considering the entire action space, (2) minimizes cross entropy between a uniform distribution over actions and the sender's estimated conditional distribution, for each communicative goal, and (3) minimizes cost of actions taken.  Importantly, the first term is derived from using the OP objective: It allows flexibility in the protocol, where multiple \textit{equivalent} actions can be mapped to the \textit{same} goal.  
The last term is derived from the costly-channel setting.  In our problem setting, since the communication channel is where cost is incurred, only sender (communicative) actions are penalized. \\

\textit{Additional Explanation.} Traversal from lines 7 to 8 follows from the application of Jenson's Inequality.  

\|*
\subsection{Task 1: \textit{No Degeneracy} -- Additional Results}
\begin{figure*}[tbh]
	\centering
	\begin{adjustbox}{minipage=\linewidth,scale=1.0}
		\centering
		\begin{subfigure}[b]{0.49\textwidth}
			\centering
			\includegraphics[width=\textwidth]{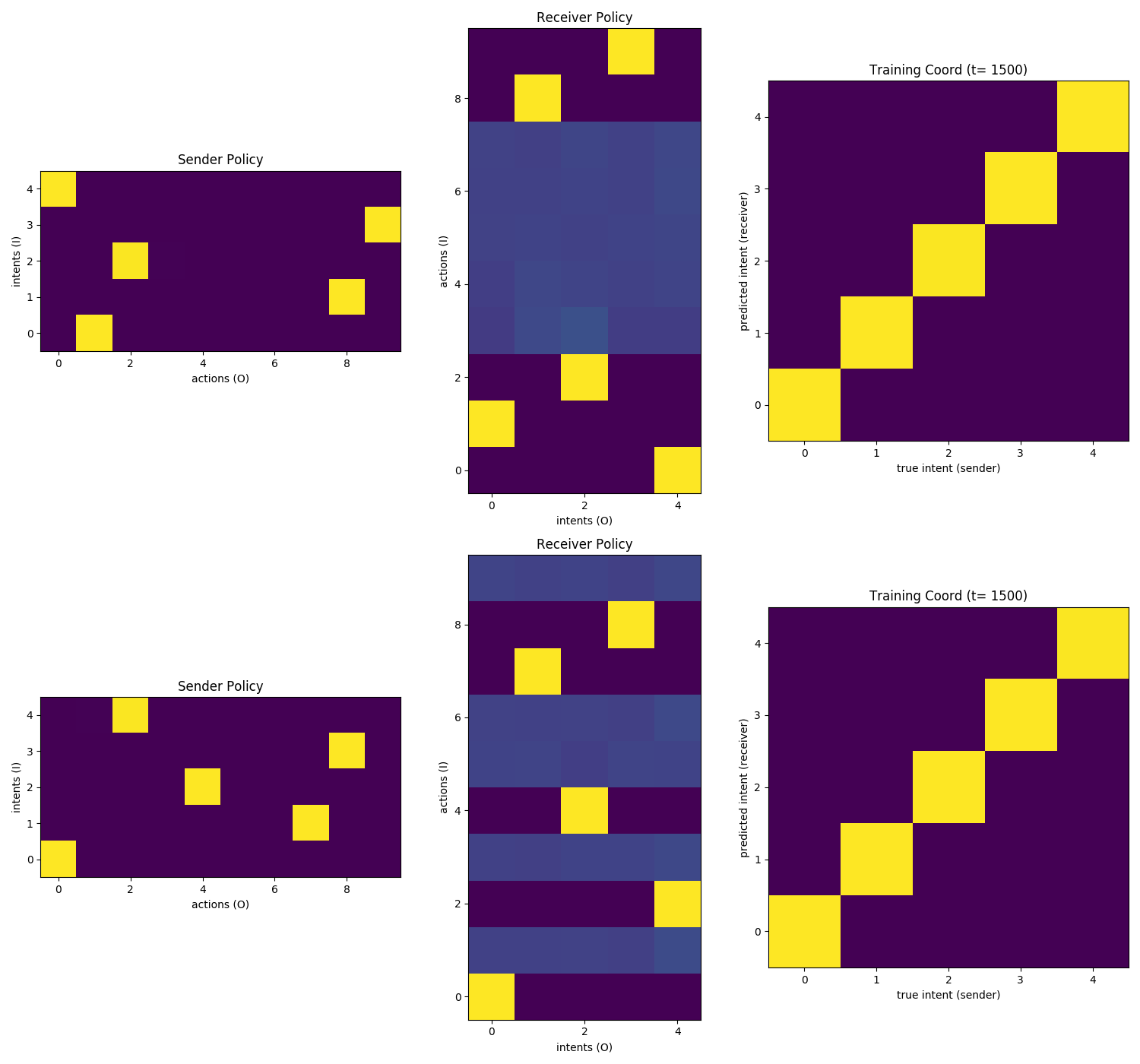}
			\caption{Zipf \textit{only} (SP)}
			\label{fig:discrete_bl_task1}
		\end{subfigure}
		\begin{subfigure}[b]{0.49\textwidth}
			\centering
			\includegraphics[width=\textwidth]{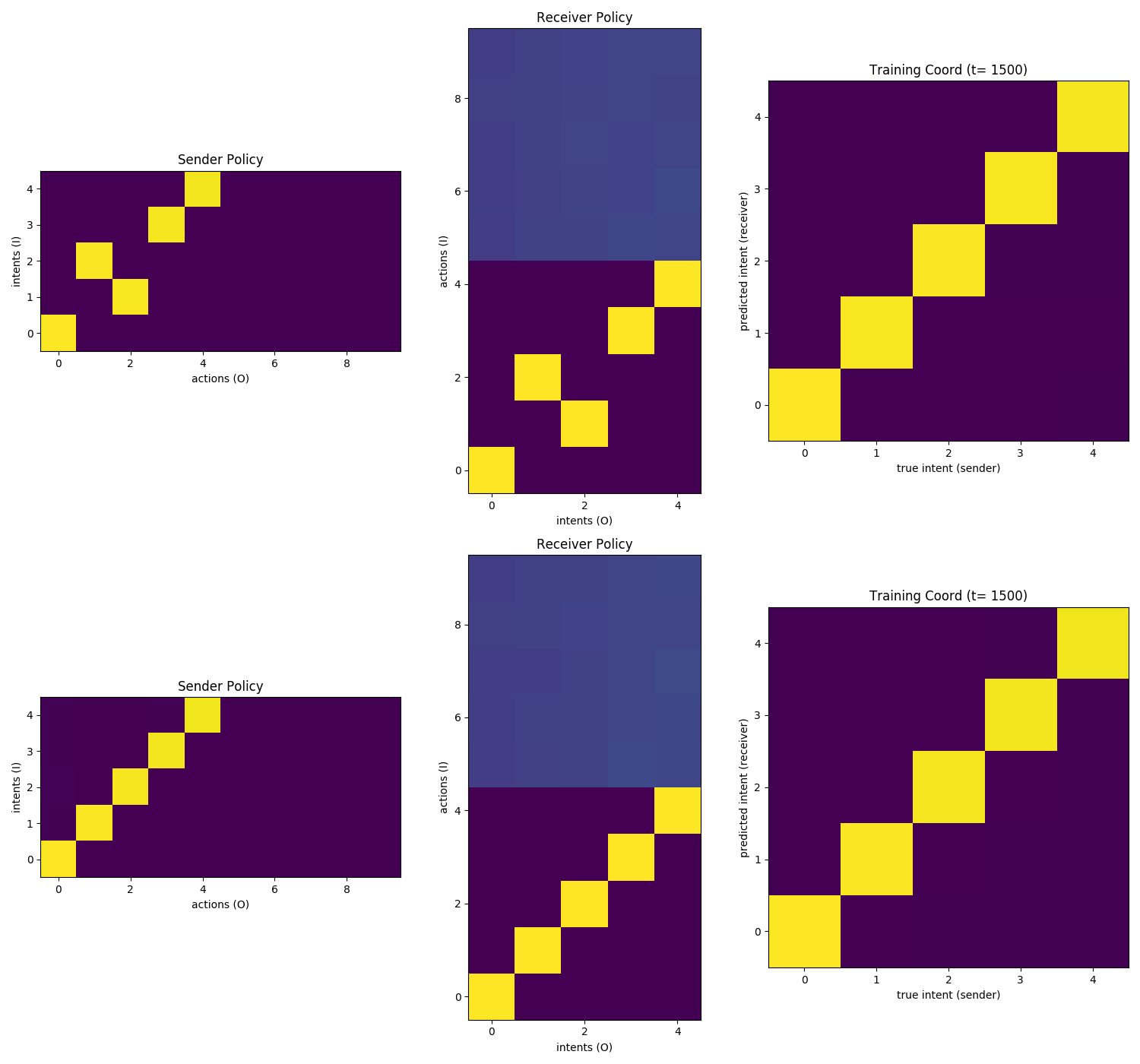}
			\caption{Zipf + Engy (SP)}
			\label{fig:discrete_ex_task1}
		\end{subfigure}
	\end{adjustbox} 
	\caption{\textbf{\textit{Example} Policies Learned from Ablation Experiment.} Protocols Learned \textit{without} (left) and \textit{with} (right) Energy Costs. \textit{No Degeneracy} Task. Zipfian Distribution over Intents. Two independently trained agent pairs (top, bottom) per condition. Left column shows sender policy mapping given intents to communication actions (messages). Middle column shows receiver policy mapping sender actions to predicted intents. Right column shows SP Performance ($p(predictedIntent \; | \; trueIntent)$) at the end of Protocol Training.  Illustrates both conditions can train sender policies to communicate effectively with training partners, but only the condition with the energy-based structure (\ref{fig:discrete_ex_task1}) produces policies sufficiently similar for generalization \textit{beyond} training partners.}
	\label{fig:ablation_experiment} 
    \vspace{-3mm}
\end{figure*}
*/

\subsection{Task 2: \textit{Energy Degeneracy} -- Additional Results}
The example policies visualized for the \textit{Energy Degeneracy} task provide some additional intuition regarding the impact of the common knowledge constraints used, on successful communication in this problem setting.

Figure \ref{fig:ablation_experiment} compares the use of cheap talk channel with the use of a costly communication channel, when trained with \ourmethod.  The latter two (Figures \ref{fig:discrete_domain_task2_bl_spcp_pairings} and \ref{fig:discrete_domain_task2_ex_spcp_pairings}) show similar qualitative analysis, but when trained using the SP baseline algorithm.  Unlike with \ourmethod, using SP, policies perform very differently when tested against training parnters (SP) versus when tested against independently trained or novel partners (XP).  So for \ourmethod, we only show ZS communication success (XP), as it is consistent with the training communication success (SP).  For the SP trained policies, we visualize training coordination (SP) and zero-shot coordination (XP) separately.

\begin{figure*}[htb]
	\centering
	\begin{adjustbox}{minipage=\linewidth,scale=1.0}
		\centering
		\begin{subfigure}[b]{0.49\textwidth}
			\centering
			\includegraphics[width=\textwidth]{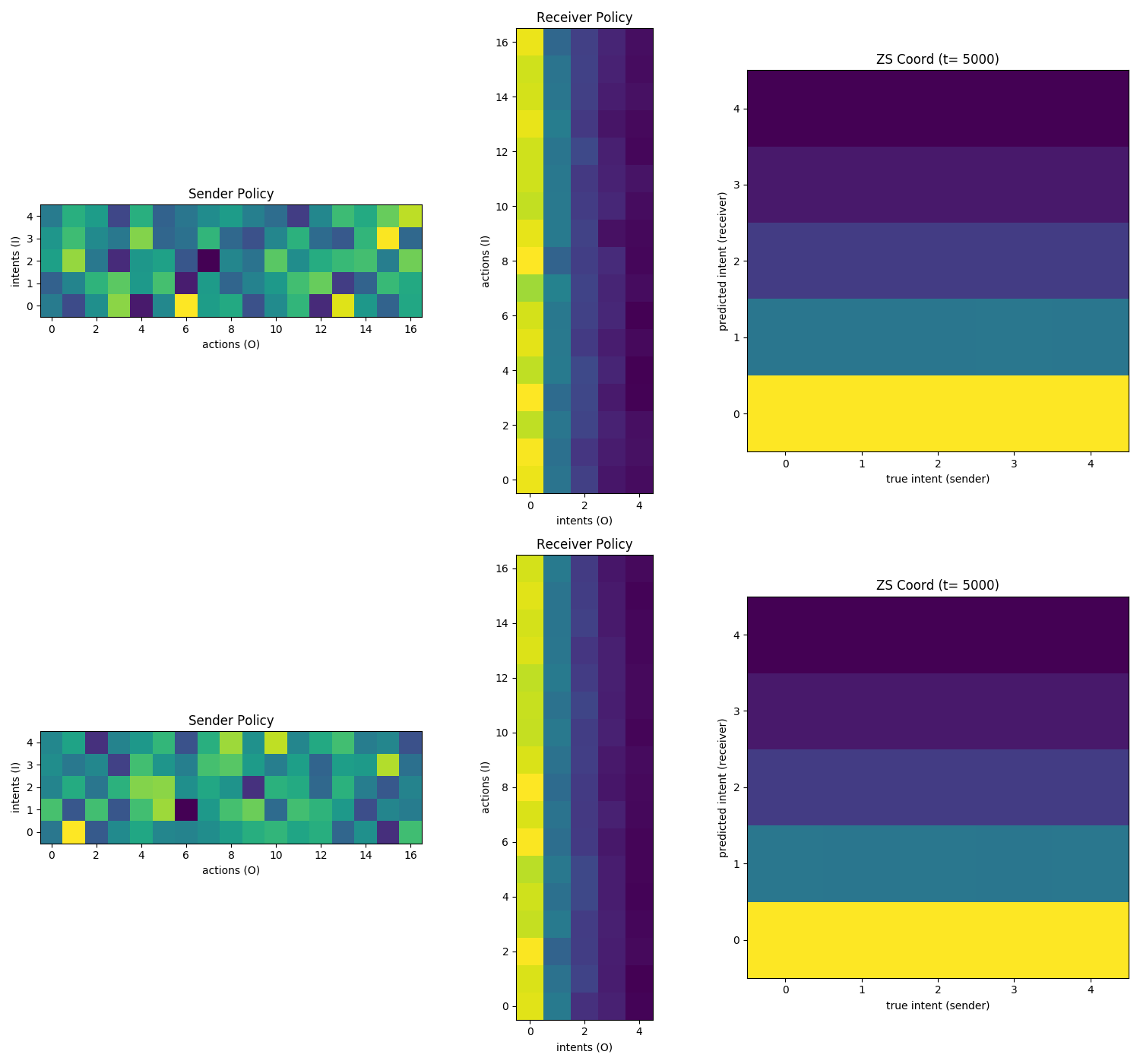}
			\caption{Zipf \textit{only} (QED): \textit{Novel} Pairings Tested [XP]}
			\label{fig:discrete_bl_task1}
		\end{subfigure}
		\begin{subfigure}[b]{0.49\textwidth}
			\centering
			\includegraphics[width=\textwidth]{figures/qual_analysis_OP-XP.png}
			\caption{Zipf + Engy (QED): \textit{Novel} Pairings Tested [XP]}
			\label{fig:discrete_ex_task1}
		\end{subfigure}
	\end{adjustbox} 
	\caption{\textbf{\textit{Example} Policies Learned from Ablation Experiment.} Compares Protocols Learned \textit{without} (left) and \textit{with} (right) Energy Costs. \textit{Energy Degeneracy} Task. Two independently trained agent pairs (top, bottom) per condition. Left column shows sender policy mapping given intents to communication actions (messages). Middle column shows receiver policy mapping sender actions to predicted intents. Right column shows SP Performance ($p(predictedIntent \; | \; trueIntent)$) at the end of Protocol Training.  All pairs of agents train with a Zipfian Distribution over Intents, for strictly ordering intents by rank, but they differ in whether they employ a cheap talk (left) or costly (right) channel, coupled with that distribution. Illustrates that using the QED algorithm, both conditions converge on policies that will perform \textit{consistently} with training partners (SP) and novel partners (XP), but only the condition with a nonuniform distribution over intents + the costly communication channel (\ref{fig:discrete_ex_task1}) produces policies that are \textit{successful} in their communication.  Intuitively, without a costly channel for mapping intents to specific actions or equivalence classes of actions, the sender policy learns to map each intent a distribution over all actions, and the best the receiver policy can do is make predictions of intent based upon likelihood of observing each intent.}
	\label{fig:ablation_experiment} 
\end{figure*}

\begin{figure*}[htb]
	\centering
	\begin{adjustbox}{minipage=\linewidth,scale=0.9}
		\centering
		\begin{subfigure}[b]{0.49\columnwidth}
			\centering
			\includegraphics[width=\textwidth]{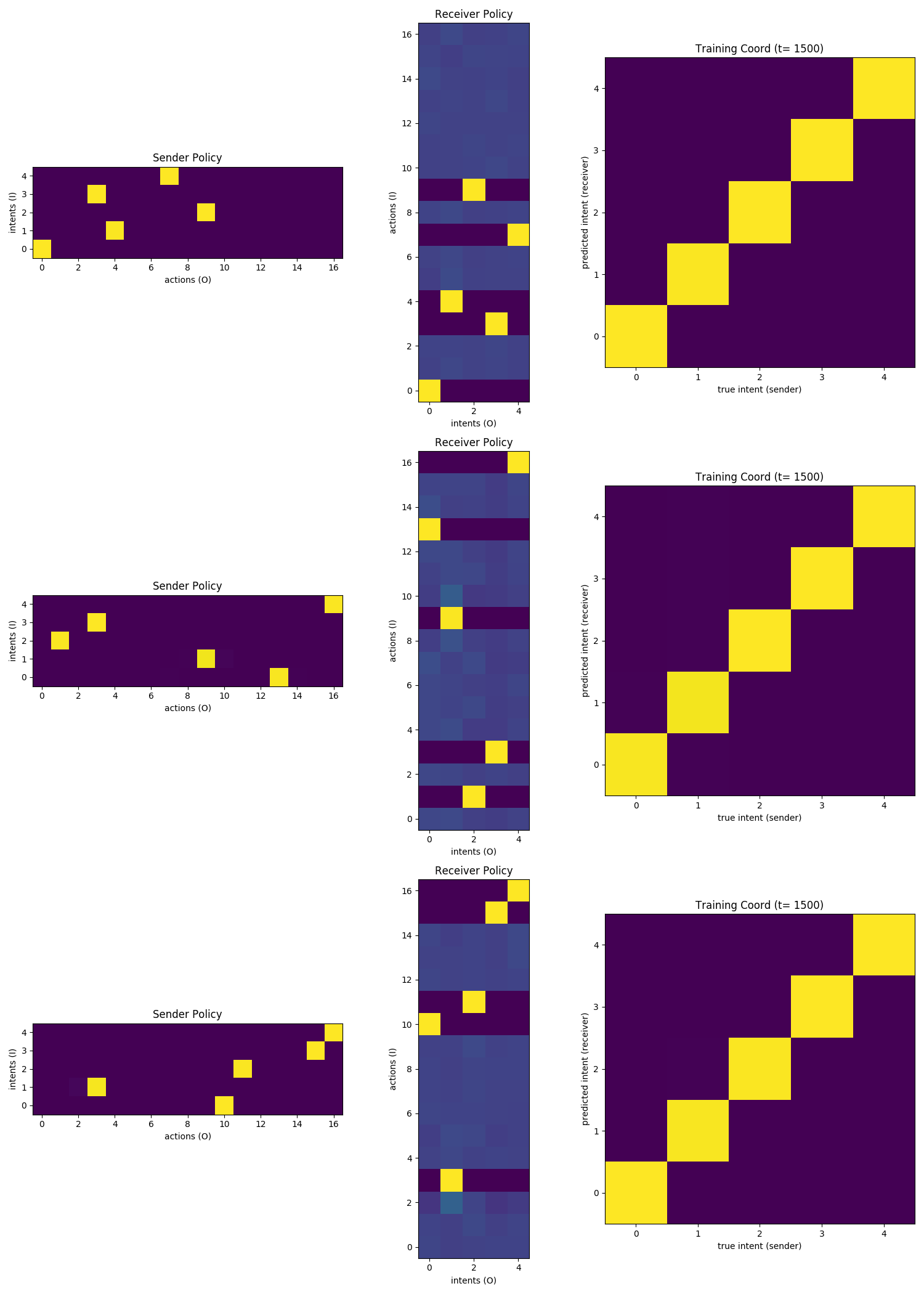}
			\caption{Zipf \textit{only} (SP): Training Pairs Tested [SP]}
			\label{fig:discrete_bl_task2_sp}
		\end{subfigure}
		\begin{subfigure}[b]{0.49\columnwidth}
			\centering
			\includegraphics[width=\textwidth]{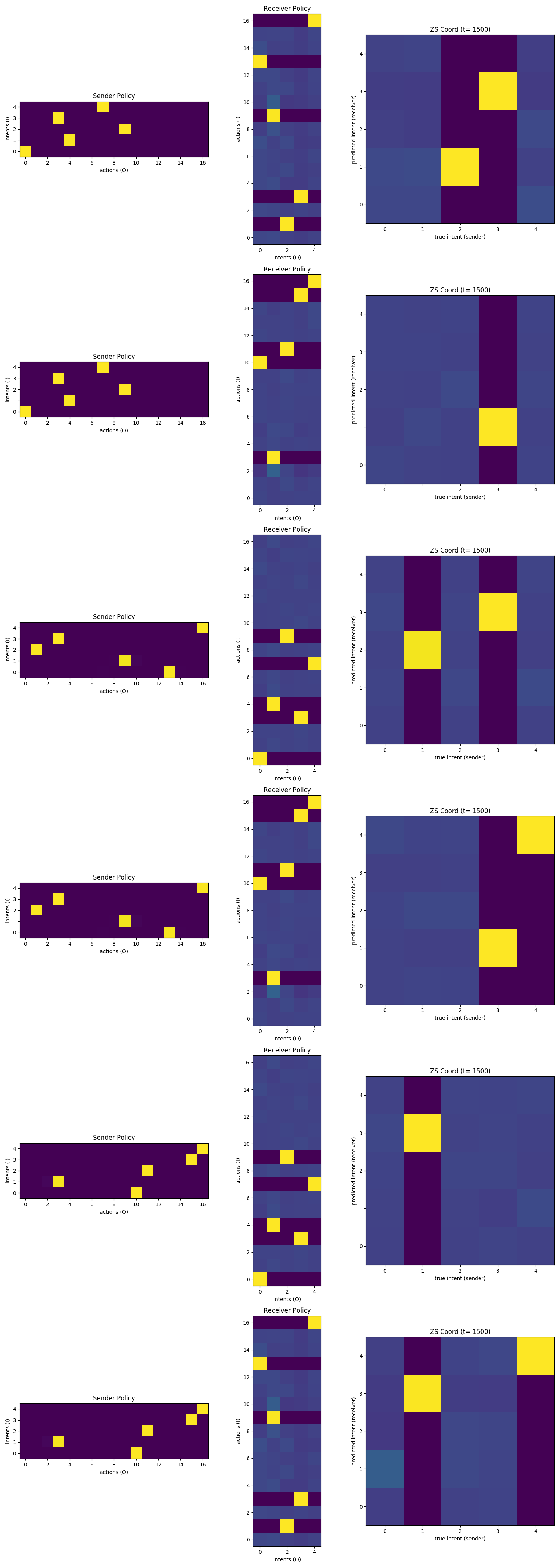}
			\caption{Zipf \textit{only} (SP): \textit{Novel} Pairings Tested [XP]}
			\label{fig:discrete_bl_task2_cp}
		\end{subfigure}
	\end{adjustbox} 
	\caption{\textbf{Protocols Learned and Tested for ZS Coordination using SP training.} \textit{Energy Degeneracy} Task. Zipfian Intent Distribution with Cheap-Talk Channel (\textit{only one} common knowledge constraint). Protocol Training Coordination (SP) and Zero-Shot Coordination (XP). Illustrates how challenging ZS communication (\ref{fig:discrete_bl_task2_cp}, right column) is as compared to communication success with training partners (\ref{fig:discrete_bl_task2_sp}, right column).  It is made harder by the fact that policies learned during the initial protocol training phase are very different, across senders (\ref{fig:discrete_bl_task2_sp}, left column).}
	\label{fig:discrete_domain_task2_bl_spcp_pairings} 
\end{figure*}

\begin{figure*}[htb]
	\centering
	\begin{adjustbox}{minipage=\linewidth,scale=0.9}
		\centering
		\begin{subfigure}[b]{0.49\columnwidth}
			\centering
			\includegraphics[width=\textwidth]{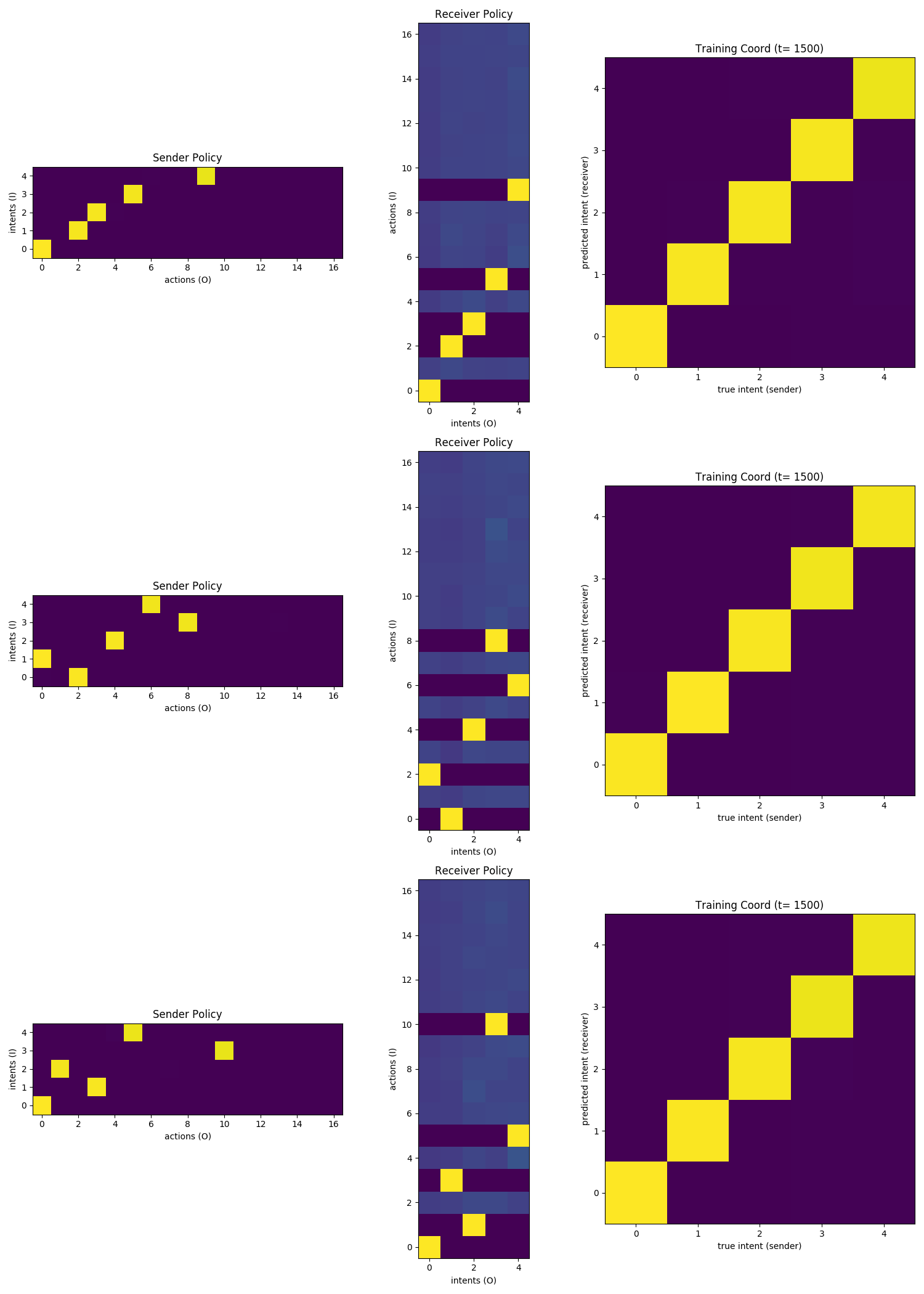}
			\caption{Zipf + Engy (SP): Training Pairs Tested [SP]}
			\label{fig:discrete_ex_task2_sp}
		\end{subfigure}
		\begin{subfigure}[b]{0.49\columnwidth}
			\centering
			\includegraphics[width=\textwidth]{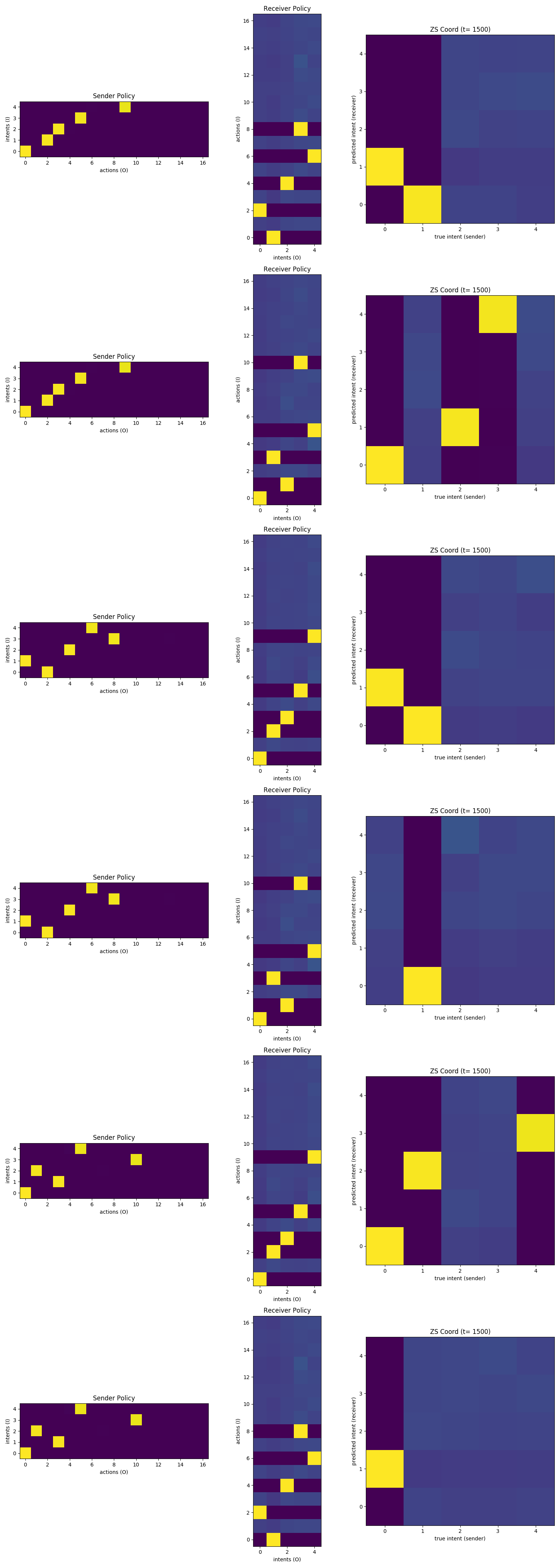}
			\caption{Zipf + Engy (SP): \textit{Novel} Pairings Tested [XP]}
			\label{fig:discrete_ex_task2_cp}
		\end{subfigure}
	\end{adjustbox} 
	\caption{\textbf{Protocols Learned and Tested for ZS Coordination using SP training.} \textit{Energy Degeneracy} Task. Zipfian Intent Distribution with \textit{Costly} Channel (\textit{both} common knowledge constraints). Protocol Training Coordination (SP) and Zero-Shot Coordination (XP).  Illustrates ZS communication (\ref{fig:discrete_ex_task2_cp}, right column) is substantially more challenging than success with training partners (\ref{fig:discrete_ex_task2_sp}, right column).  Nonetheless, ZS coordination from protocols built upon energy-based latent structure is slightly more successful as compared against ZS coordination in baseline condition with no energy penalty (\ref{fig:discrete_ex_task2_cp}, right column).  The trends observed are similar to that of comparing these two conditions, using QED.}
	\label{fig:discrete_domain_task2_ex_spcp_pairings} 
\end{figure*}

\fi
 


\end{document}